\documentclass[aip, jcp, url, amsmath, amssymb,
 longbibliography, nofootinbib,
reprint,
]{revtex4-1}

\usepackage{graphicx}% Include figure files
\usepackage{dcolumn}% Align table columns on decimal point
\usepackage{bm}% bold math
\usepackage{hyperref} % make the links clickable
\usepackage[capitalize]{cleveref}
\usepackage{soul, color}
\begin{document}

\title{Perspective: Markov Models for Long-Timescale Biomolecular Dynamics}
% in my mind, MSM encapsulates a lot, not just the Markovian part, but also
% 1) automatic state decomposition
% 2) internal checks of model quality
% I want to make sure that these key aspects aren't lost, so that's why I
% suggest put MSM in the title.

\author{C. R. Schwantes}
\thanks{These two authors contributed equally}
\affiliation{Department of Chemistry, Stanford University, Stanford, CA 94305}
\author{R. T. McGibbon}
\thanks{These two authors contributed equally}
\affiliation{Department of Chemistry, Stanford University, Stanford, CA 94305}
\author{V. S. Pande}
\affiliation{Department of Chemistry, Stanford University, Stanford, CA 94305}
\affiliation{Department of Computer Science, Stanford University, Stanford, CA
94305}
\affiliation{Department of Structural Biology, Stanford University, Stanford,
CA 94305}
\affiliation{Biophysics Program, Stanford University, Stanford, CA 94305}
\date{\today}

\begin{abstract}
Molecular dynamics simulations have the potential to provide atomic-level
detail and insight to important questions in chemical physics that cannot be
observed in typical experiments. However, simply generating a long trajectory
is insufficient, as researchers must be able to transform the data in
a simulation trajectory into specific scientific insights. Although this
analysis step has often been taken for granted, it deserves further attention
as large-scale simulations become increasingly routine. In this perspective, we
discuss the application of Markov models to the analysis of large-scale
biomolecular simulations. We draw attention to recent improvements in the
construction of these models as well as several important open issues. In
addition, we highlight recent theoretical advances that pave the way for a new generation of models of molecular kinetics.
\end{abstract}

%\pacs{Valid PACS appear here}
\maketitle

\section{Introduction}

Simulations have the hope to shed light on many areas in chemical physics.
For example, molecular dynamics (MD) simulations, which numerically integrate
Newton's equations to simulate physical dynamics, can
provide high-resolution physics-based models of emergent
biological phenomena.\cite{ivetac2009elucidating, lindorff2011fast,
ostmeyer2013recovery, shukla2014activation}
While promising, there are three central challenges which limit the
application of MD for studying questions in biophysics. The first is the
development of simplified physical models (called force fields), which avoid the
intractability of solving the full, electronic Schr{\"o}dinger
equation. Second, while simulations are required to use a discrete
time step on the order of one femtosecond, processes that are characterized by
slow, large-scale, collective motions, such as protein folding, protein-ligand
binding, and conformational change can take milliseconds or longer to occur. 
This separation in timescales requires a simulation of length at least $10^{12}$ 
time steps in order to observe one such event, which is difficult using
current hardware. Finally, the analysis of
MD simulations is itself non-trivial since the result -- a set of trajectories
tracking the Cartesian coordinates of every particle -- can contain millions of
data points in tens of thousands of dimensions.
Despite these challenges, a number of recent innovations
have expanded the scope of molecular simulation.
In fact, because of improvements in forcefield accuracy,\cite{piana2011robust,
lindorff2010improved,
  wang2013systematic, wang2014building, lamoureux2003simple,
  shi2013polarizable}
and the development of novel computing platforms,\cite{shirts2000screen,
  shaw2008anton, eastman2013openmm, gotz2012Routine, pronk2013gromacs}
we believe that quantitative analysis has increasingly become a limiting factor in the
application of MD.\cite{freddolino2010challenges,
  lane2013milliseconds}

With routine MD datasets now comprising terabytes of data, the direct
visualization of raw MD trajectories is neither scalable nor quantitative.
Instead, we suggest that MD trajectories should not be viewed as ends in and of
themselves, but instead a means of parameterizing a quantitative statistical
model of the structure and dynamics of the system of interest.

These models should have the following properties:
\begin{enumerate}
\item The model should be oriented toward quantitatively describing the
long-timescale processes in the data.
\item The model should be \emph{suitably complex}, capable of smoothly
  adapting to -- rather than assuming -- the structure of the data.
\item The model should be interpretable and provide answers to specific
scientific questions that may be difficult to answer via experiment alone.
\end{enumerate}
Towards this end, we survey recent developments in the theory and
application of Markov modeling for the analysis of MD simulations, which
have all of these desired properties.

\section{Long Timescales and the Transfer Operator}

The transfer operator provides a theoretical
foundation for the description of the slow dynamical processes in
stochastic systems, and is the basis for the approaches discussed herein. Indeed,
at its essence, the methods described below all work to build some sort of
numerical model of the transfer operator.  Here we highlight
some important details; we refer the interested reader to Sch\"utte, Huisinga, 
and Deuflhard. \cite{schutte2001transfer}

Consider a time-homogeneous, ergodic Markov
processes, $\mathbf{x}_t$, in phase space, $\Omega$, which is reversible with
respect to a positive stationary density $\mu(\mathbf{x})$.
At time $t$, an ensemble of such processes can be described
by a probability distribution, $p_t(\mathbf{x})$, which can change as a
function of time. The evolution of this probability density over a time
interval $\tau$ can be described by the action of an operator, $\mathcal{T}(\tau)$,
the transfer operator. For each time $t$, define the equilibrium-weighted
probability density, $u_t(\mathbf{x}) = p_t(\mathbf{x})\mu(\mathbf{x})^{-1}$.
Then $\mathcal{T}(\tau)$ is defined as the operator that evolves $u_t(\mathbf{x})$ to
$u_{t+\tau}(\mathbf{x})$:
\begin{equation}
u_{t+\tau}(\mathbf{x}) = \mathcal{T}(\tau) \circ u_t(\mathbf{x}).
\end{equation}
$\mathcal{T}(\tau)$ is bounded and self-adjoint with respect to the scalar product
\begin{equation}
\langle f, g \rangle_\mu = \int_\Omega d\mathbf{x} \, f(\mathbf{x})
g(\mathbf{x}) \mu(\mathbf{x}),
\end{equation}
and can therefore be decomposed into terms corresponding to each of its
eigenfunctions:
\begin{equation}
\label{eq:transfer_eigs}
u_{t + n\tau}(\mathbf{x}) = \big[\mathcal{T}(\tau)\big]^n \circ
u_t(\mathbf{x}) = \sum_{i=1}^\infty \lambda_i^n \langle \psi_i, u_t
\rangle_\mu \psi_i(\mathbf{x}),
\end{equation}
The eigenvalues, $\lambda_i$, are real, and the eigenpairs ($\lambda_i, \psi_i(\mathbf{x})$) can be taken to be sorted in decreasing order by eigenvalue. According
to the Perron-Frobenius theorem, $\lambda_1$ is equal to one and corresponds to
the stationary eigenfunction, $\psi_1(\mathbf{x})=1$. The remaining eigenvalues lie in
$(-1,1)$, and are associated with eigenfunctions which describe directions of
collective motion in phase space -- dynamical processes -- along which the
system relaxes towards equilibrium. \cref{eq:transfer_eigs} can therefore be
rewritten as:
\begin{equation}
\label{eq:transfer_eigs2}
u_{t + n\tau}(\mathbf{x}) = 1 + \sum_{i=2}^\infty \exp \left(-\frac{n\tau}{t_i}\right) \langle \psi_i, u_t
\rangle_\mu \psi_i(\mathbf{x}). 
\end{equation}
where $t_i$ are the characteristic relaxation timescales of each dynamical mode, defined in terms of the associated eigenvalues,
\begin{equation}
t_i = - \frac{\tau}{\ln \lambda_i}.
\end{equation}
At long times (large $n$), terms in \cref{eq:transfer_eigs} and \cref{eq:transfer_eigs2}
corresponding to large eigenvalues (long relaxation timescales) dominate the summation,
as small-eigenvalue modes equilibrate and decay quickly. The focus for accurate numerical
models of $\mathcal{T}(\tau)$ is thus to resolve these dominant eigenmodes.

The fundamental approach we will take is to use computational methods to build
an approximate representation of the transfer operator, $\mathcal{T}(\tau)$, from
molecular dynamics simulations in a way which is systematic, automated, and
statistically rigorous.  One major upshot of this approach is that it allows a
wide range of simulation data to be useful, ranging from many relatively short
trajectories (compared to the longest timescales of interest) to a few long
trajectories.  Below, we lay out the challenges, current status, and future of
this paradigm.

\section{Methods}

\subsection{Markov State Models}

Markov State Models (MSMs) describe the stochastic dynamics of
molecules as a Markovian jump process between a finite number of
conformational states. Because of the Markov property, the probability
of jumping to a new state only depends on the current state.
As such, the dynamics are completely determined by a transition matrix,
$\mathbf{T}$, such that
\begin{equation}
T_{ij} = P(\mathbf{x}_{t+\tau} \in s_j | \mathbf{x}_t \in s_i)
\end{equation}
where $\{\mathbf{x}_t\}_{t=1}^N$ is a trajectory in $\Omega$ and the
states, $S = \{s_i\}_{i=1}^k$, are a set of non-overlapping subsets of
$\Omega$ that partition the space: $s_i \subseteq \Omega \;\forall\; i$, $\cup_{i=1}^k s_i = \Omega$, $s_i \cap s_j = \emptyset \;\forall\; i \neq j$. $\mathbf{T}$ is a discretization of $\mathcal{T}$,
\begin{equation}
\label{eq:disc_t}
T_{ij} =\frac{\int_{s_j} d\mathbf{x} \, \mu(\mathbf{x}) (\mathcal{T} \circ
\mathbf{1}_{s_i})}{\int_{s_i} d\mathbf{x} \, \mu(\mathbf{x})},
\end{equation} where
\begin{equation}
\mathbf{1}_{s_i}(\mathbf{x}) = \left\{ \begin{array}{cc} 1 & \mathbf{x} \in s_i
\\ 0 & \mathbf{x} \notin s_i \end{array} \right.
\end{equation}

There is a rich history of describing the dynamics of biophysical
processes in terms of finite state models, including work based on
enumeration of potential energy minima,\cite{czerminski1989reaction,
  evans2004folding} study of simplified physical
models,\cite{grubmuller1994molecular, scala2001small} analysis of
long-timescale processes from many short
trajectories,\cite{rao2004protein, faradijan2004computing,
  andrec2005protein, chodera2006long, buchete2008coarse,
  pan2008building, noe2009constructing} and transition path and
spectral analysis of the resulting network
models.\cite{berezhkovskii2009reactive, e2006towards,
  bowman2012improved, deuflhard2000identification} For more specific details on
MSMs, we refer the interested reader to \citet{prinz2011markov} and Pande,
Beauchamp, and Bowman.\cite{pande2010everything}

Though the theory is relatively simple, the construction of an MSM is not
a solved problem. Given a set of MD simulations, we need to define a state
decomposition, $S$, as well as estimate the transition probability matrix,
$\mathbf{T}$.
Straightforward maximum likelihood estimators for $\mathbf{T}$, given $S$ exist,
\cite{prinz2011markov, beauchamp2011msmbuilder2} but the determination of $S$ itself is more challenging. A wide variety of strategies
have been proposed for constructing $S$, including grid-based
discretization,\cite{buchete2008coarse} and clustering with many
algorithms\cite{beauchamp2011msmbuilder2, senne2012emma} and distance
metrics.\cite{mu2005energy, kellogg2012evaluation,
  zhou2012distribution, mcgibbon2013learning, perez2013identification,
  schwantes2013improvements} However, no consensus has emerged in the literature on
the preferred strategy, largely because these clustering schemes rest on
heuristic foundations built primarily by physical intuition.
No solid theoretical justifications exist for choosing one of these schemes over
another, especially when statistical errors in the parameterization of $\mathbf{T}$
are non-negligable.

On the other hand, \citet{sarich2010approximation} showed that in the absence of
statistical uncertainty in the parameterization of $\mathbf{T}$, an MSM's error,
measured in terms of the maximum possible difference between the true and
MSM-predicted probability density at a later time, is bounded. The state
decomposition-dependent component of this bound quantifies the difference between
the transfer operator's continuous eigenfunctions, $\psi_i(\mathbf{x})$, and the
MSM's piecewise constant approximate eigenfunctions, given by the eigenvectors of
$\mathbf{T}$. This result implies that a
``good'' state decomposition is one that finely partitions phase space particularly in
the regions where the dominant eigenfunctions, $\psi_i(\mathbf{x})$, change rapidly,
such as transition regions or energy barriers. Unfortunately, these barrier regions are
precisely the areas that are most poorly sampled in MD. Finely subdividing these regions
into many MSM states can lead to large statistical errors in the associated transition probabilities.

In practice, even the seemingly simple problem of deciding the correct number of states
given a clustering algorithm and distance metric is
difficult.\cite{kellogg2012evaluation, mcgibbon2014statistical}
When the number of states is low, the discretization in \cref{eq:disc_t}
is very coarse, which leads to systematic underestimation of the relaxation
timescales.\cite{sarich2010approximation, djurdjevac2012estimating} However, as
the number of states increases, we have to estimate more state to state
transition probabilities. With the amount of data fixed, increasing the number
of parameters leads to a corresponding increase in the statistical uncertainty
of the model. Some recent work, however, has made progress toward balancing these
sources of errors by defining a likelihood function for evaluating and comparing
models.\cite{kellogg2012evaluation, mcgibbon2014statistical} These likelihood
functions are useful for simple systems, but can be difficult to employ
in most MSM applications.

To illustrate the difficulty of building a state space more concretely,
consider a simulation of a protein and a ligand where we
are interested in calculating the on and off rates of the binding reaction. In
this simulation, the ligand is very mobile when it is not bound to the protein,
and so a typical geometric clustering will produce many unbound states. These
states contribute to the statistical error since we are essentially forcing the
model to describe the diffusion of the ligand in water. However, since we are
only interested in the binding reaction, we do not care about this diffusion,
and could instead build a more statistically robust MSM by ignoring these
motions in the unbound state. This behavior is not unique to protein-ligand
simulations, and in fact has been an issue
when analyzing protein folding simulations. Typically these models must use
tens of thousands of small states -- termed microstates -- in order to accurately
describe the folding reaction's
timescale.\cite{bowman2010atomistic, lane2011markov, voelz2012slow}
Not only does this increase the statistical error, but it raises major 
challenges for the interpretation of the resulting model. One suggested 
resolution has been to lump these microstate 
models into coarser macrostate models by grouping together states which 
rapidly interconvert. \cite{deuflhard2000identification, deuflhard2005161, 
	bowman2012improved,	bacallado2009bayesian}

In our view, this microstate-to-macrostate lumping strategy can be dangerous, and we
discourage its use. The lumping procedure relies directly on the quality of the
microstate model, which is purposely estimated subject to large statistical errors.
For example, the PCCA and PCCA+ algorithms for lumping MSMs are designed
to preserve the estimates of the slow eigenfunctions \emph{as identified by the
microstate model},\cite{deuflhard2000identification, deuflhard2005161} without regard
to the fact that these estimates are, by design, approximated subject to large errors.
Although other methods like BACE partially remedy this
problem,\cite{bowman2012improved, bowman2013quantitative} in our view a more direct solution is to construct accurate and interpretable models from the start.

To be able to build simultaneously accurate and interpretable MSMs,
clustering must be able to focus on the important (slowly equilibrating)
degrees of freedom, while ignoring the unimportant ones. One suitable method is
thus to
perform dimensionality reduction before clustering, explicitly removing quickly
decorrelating coordinates. With this in mind, \citet{schwantes2013improvements}
and \citet{perez2013identification} introduced time-structure based
Independent Component Analysis (tICA) for building MSMs. The tICA
method identifies the slowest decorrelating linear projections in the
observed data. This technique has made it possible to build macrostate MSMs
from the very beginning that better resolve long-timescale processes.
\cite{mcgibbon2014statistical, lapidus2014complex} In addition, these models have
identified new slow motions that eluded detection previously, such as
near-native register shift dynamics in $\beta$-sheet
proteins.\cite{schwantes2013improvements}

\subsection{A Generalization of MSMs: Hidden Markov Models}

Part of the difficulty in constructing the MSM state space is that there is no
single quantitative criterion for comparing alternative state spaces which is
suitable for practical applications. Therefore, one approach is to generalize
the MSM method such that the state decomposition can be optimized
simultaneously with the transition probabilities in a unified way. Hidden
Markov models (HMMs) are one such extension that relax the
constraint that states correspond to a discrete partition of $\Omega$.
Like MSMs, HMMs characterize the system using a Markov jump process over a finite
number of states; however in the HMM, these states are not assumed to be
directly observed. Instead, each hidden state is equipped with an emission
distribution defined on $\Omega$, describing the per-state conditional
probability of observing a conformation at a particular point in phase space.
HMMs have been widely used in a variety of fields, including signal processing,
speech, and bioinformatics.\cite{rabiner1989tutorial, stanke2003gene}

For example, using Gaussian emissions, if $\{\mathcal{S}_t \in \{1\ldots
k\}\}_{t=1}^N$ is the Markov chain over hidden states with transition
probability matrix $\mathbf{T}$, $\{X_t \in \Omega\}_{t=1}^N$ is the
observed process, $\pi$ is a probability vector for the initial
distribution over hidden states, $\{\mu_i \in \Omega\}_{i=1}^{k}$ are
the per-state means, and $\{\Sigma_i\}_{i=1}^k$ are the per-state
covariance matrices, the model can be specified probabilistically as
\begin{align}
\mathcal{S}_1 &\sim \mathrm{Categorical}(\pi), \\
\mathcal{S}_{t+1} &\sim \mathrm{Categorical}(\mathrm{row}_{S_t}(\mathbf{T})), \\
X_t &\sim \mathcal{N}(\mu_{\mathcal{S}_t}, \Sigma_{\mathcal{S}_t}).
\end{align}

Constructing an HMM requires estimating the transition probability matrix
$\mathbf{T}$ as well as the emission distribution parameters $\theta_i
= (\mu_i$, $\Sigma_i)$. Using a maximum-likelihood approach,
$\mathbf{T}$ and $\theta_i$ can be estimated
jointly.\cite{dempster1977maximum} In the HMM, the $\theta_i$'s are
analogous to the MSM's state decomposition, but unlike the MSM state
space, the $\theta_i$'s are determined simultaneously with the elements
of $\mathbf{T}$ -- the two are jointly optimized with respect to the \emph{same} objective function. In our view, this is the key advantage of the HMM, as
the optimal construction of the MSM state space remains an unsolved
problem.

One caveat in applications of HMMs to analysis of MD is that without
conditioning on $\mathcal{S}_t$, the HMM's observed process $X_t$ is
strictly non-Markovian when the emission distributions have nonzero
overlap. While it is precisely these overlaps that make the model
(particularly $\theta_i$) optimizable, they also hamper direct
interpretation of the HMM as a model for the Markovian transfer
operator.\cite{noe2013projected, mcgibbon2014understanding}

\section{Recent Applications}

Markov models have been heavily applied to many different types of systems,
ranging from protein folding,\cite{lane2011markov, voelz2012slow,
baiz2014molecular} RNA folding,\cite{huang2009constructing} protein-ligand
binding,\cite{buch2011complete}
and protein conformational change.\cite{shukla2014activation,
mcgibbon2014understanding, sadiq2012kinetic}
Below, we summarize two recent highlights from the literature, which
demonstrate the unique utility of Markov modeling. 
In particular, these approaches allow researchers to connect
many short MD trajectories into a single coherent description of the
dynamics. Moreover, MSMs are able to provide an interpretable
picture of the kinetics that can lead to specific scientific insight.

\subsection{$\beta_2$ Adrenergic Receptor ($\beta_2$AR) Activation}
$\beta_2$AR is a G-protein coupled receptor (GPCR) involved in many
trans-membrane signaling pathways. GPCRs are critical in medicine,
representing the target for 30\% of top-selling drugs on the
market.\cite{chalmers2002use} Despite extensive experimental
studies,\cite{rosenbaum2007gpcr,cherezov2007high, laporte1999beta2,
  jordan2001oligomerization} the mechanism of $\beta_2$AR activation
is elusive. \citet{kohlhoff2014cloud} used Google exacycle, a cloud
computing platform that enables scientific calculations to be run on
Google's datacenter infrastructure, to collect over 2 ms of aggregate
MD simulations of $\beta_2$AR in a lipid bilayer and explicit solvent
using tens of thousands of individual MD trajectories, each with a mean length of
$\sim 10$ ns.
Although no individual MD trajectory traversed the full activation pathway,
\citet{kohlhoff2014cloud} were
able to use MSMs to understand the full activation landscape using
these many short trajectories. This is a particularly powerful
property of MSMs, because collecting many short trajectories is
substantially easier than generating a few long trajectories.\cite{buch2010high,
  shirts2000screen}

In addition, \citet{kohlhoff2014cloud} simulated these trajectories in
three waves. The starting conformations for the second and third wave
were selected by uniformly sampling states from an MSM built on the
previous waves. Known as adaptive sampling, this technique provides a framework
for efficiently covering all of phase space with many, relatively short
simulations, which is only possible because of the use of
MSMs.\cite{singhal2005error, weber2011characterization}

\citet{kohlhoff2014cloud} used the MSM to show
that the activation of $\beta_2$AR proceeded along multiple paths. Many of these
pathways featured metastable intermediates that displayed unique
chemotype selectivity in docking studies, suggesting that
computational drug design can be enhanced by considering important
intermediates.

\subsection{$\beta$-Lactamase Allosteric Sites}

While typical structure-based computational drug design pipelines use a
single protein conformation to screen for potential
therapeutics,\cite{kalyaanamoorthy2011structure}
proteins are dynamic systems and an ensemble of structures exist in solution.
Accounting for this heterogeneity could expand the range of possible methods
for modulating protein activity. With this in mind,
\citet{bowman2012equilibrium} set out to
find cryptic druggable binding sites in $\beta$-lactamase using MD
simulations and MSMs.

Using one hundred microseconds of aggregate simulation time
distributed over hundreds of trajectories, each no longer
than 500 ns, \citet{bowman2012equilibrium} were able to characterize slow
near-native conformational changes in $\beta$-lactamase. Their MSM
was able to describe the kinetics of the opening and closing of various
cryptic binding sites on distal regions of the
protein. In particular, MSMs allowed the researchers to study the
lifetimes and equilibrium opening probabilities of these pockets. They
first observed that a known cryptic binding site was predicted to be
open 53\% of the time according to the MSM. Furthermore, they
identified 50 novel sites, which were predicted to be open more than
10\% of the time, suggesting new avenues for drug design.

\section{Discussion}

While these results highlight that MSMs have been useful for analyzing
MD simulations, the definition of the state space remains a substantial hurdle in MSM construction.
Although recent work has provided significant improvements in this area, one way to side-step
this challenge is to build a Markov model (i.e. provide an estimate 
of the transfer operator) \emph{without using a discrete state decomposition}.
Towards this end, \citet{noe2013variational} and \citet{nuske2014variational} established
a variational principle that applies to the estimation of the transfer
operator's eigenspectrum. This is analogous to the variational principle for
the Hamiltonian in quantum mechanics and guarantees that, when the operator
is discretized using a finite basis set, its eigenvalues are underestimated
in the limit of vanishing statistical errors.

Given a basis set, the variational approach searches for linear combinations
of basis set elements which approximate the transfer operator eigenfunctions
by solving a particular generalized eigenvalue problem, or equivalently maximizing a series
Rayleigh quotients that can be interpreted as autocorrelation functions. In fact, the MSM
approach is a specific instance of this method when the basis set consists of indicator
functions which partition $\Omega$, but the method more broadly can be applied
to any set of basis functions including those which have nonzero overlap.
Furthermore, tICA is also an instance of this approach, with a basis set instead
consisting of linear functions of the input coordinates. In the HMM, however,
the
process observed in $\Omega$ is not strictly Markovian, and thus its
interpretation in terms of the Markovian transfer operator (or its
eigenspectrum) is less straightforward.

This theoretical advance suggests a tension between two distinct but related
perspectives for building simplified models of molecular kinetics:
\begin{enumerate}
\item Parameterize a probability distribution over trajectories.
\item Find the first $m$ eigenfunctions of $\mathcal{T}$.
\end{enumerate}

The HMM and tICA methods are certainly different, but to what
extent are probabilistically- and variationally-inspired models distinct?
These views are, at least, not always different since an MSM is both a
probabilistic
model and one that estimates the slowest eigenfunctions of $\mathcal{T}$.
But, outside of MSMs, when is a probabilistic model more desirable than a
variational model, or vice
versa? Is one model missing something that can only be found in the other?
If so, is it possible to unify these views with a model other than an MSM? 
We do not know the answers to these questions, but both perspectives
have enriched our understanding of molecular kinetics, and we
anticipate future developments will clarify the apparent tension and lead to
new classes of Markov models.

\section{Conclusions}

The future of Markov modeling is bright. Ten years ago, conventional
MD simulations could only access timescales in the
nanosecond regime.\cite{karplus2002molecular} A revolution in GPU
computing,\cite{friedrichs2009accelerating, buch2010high, eastman2013openmm}
however, has enabled researchers around the world to
perform simulations that were once impossible. But, MD trajectories alone are
not enough. Quantitative analysis, which
can turn a simulation into scientific insight, is a necessary component
of the research process that has often been taken for granted.

Indeed, the {\em combination} of improvements in simulation and analysis
has allowed for simulations of phenomena on the hundreds of microseconds to
millisecond timescale;\cite{lane2013milliseconds}
as a high end GPU can today simulate over 100 ns/day for a 30,000 atom system, a
cluster of 100 GPUs can simulate an aggregate of 10 $\mu$s/day, yielding a
millisecond of aggregate simulation in three months.  While this cluster cannot
simulate one long millisecond trajectory in three months, MSMs allow the use of these 10-$\mu$s trajectories to describe phenomenon on the millisecond timescale.

Markov modeling is a technology that has been rapidly maturing for MD analysis.
At least two open-source software packages allow non-experts to construct
these models routinely,\cite{beauchamp2011msmbuilder2, senne2012emma} and 
tutorials on their use are available.\cite{bowman2014tutorial}
Nevertheless, many challenges remain in their theory and application.
In particular, parameter selection continues to be difficult
-- e.g. how fine a discretization best balances competing sources
of error?
Recent work has benefited from a synthesis of concepts from machine learning
and mathematical physics, and we anticipate that the future development of Markov
modeling will continue to
be enhanced by a cross-pollination of ideas from these communities.

Looking to the next 10 years, we expect that these approaches will become more
automated, with statistical methods replacing empirical tests of model quality
and adaptive sampling approaches maturing to the point where they can
automatically tackle a wide range of sampling challenges.  Moreover, in 10
years, GPUs (which double in speed every year) will likely be 1,000 times more
powerful.  The combination of these technologies suggests that MSM simulations
on the millisecond timescale will become routine and require only a day on
a modest 10-GPU cluster, and simulations on the seconds timescale will become
within reach.  At that point, the field of bimolecular simulation will likely
be in a golden period, where sampling is no longer the principle challenge for
many biological questions of interest.

\section{Acknowledgements}

The authors would like to acknowledge Thomas J. Lane for useful discussions during the 
preparation of this manuscript. In addition, the authors acknowledge funding 
from NSF-MCB-0954714 and NIH-R01-GM062868.

%\bibliography{bibliography}


\end{document}